\documentclass{eas}
\usepackage{graphicx}
\usepackage{amssymb}
\usepackage{astron}
\def\araa{ARA\&A}%
\def\apj{ApJ}%
\def\apjl{ApJ}%
\def\aap{A\&A}%
\def\mnras{MNRAS}%
\begin{document}

\title{Evolution of the first stellar generations} 
\runningtitle{Hirschi \etal: First stellar generations}
\author{R. Hirschi}
\address{Physics and Astronomy, University of Basel, Klingelberstr. 82, 4056 Basel, Switzerland}
\email{raphael.hirschi@unibas.ch}
\author{A. Maeder}\address{Geneva Observatory, Ch. des Maillettes 51, 1290 Sauverny, Switzerland}
\author{G. Meynet}\sameaddress{2}
\author{C. Chiappini}\sameaddress{2}
\secondaddress{Osservatorio Astronomico di Trieste, Via G. B. Tiepolo 11, I - 34131 Trieste, Italia}%
\author{S. Ekstr\"om}\sameaddress{2}
\begin{abstract}
Although the theoretical study of very low metallicity ($Z$) and metal--free stars is not new, 
 their importance has recently greatly increased since two related 
fields have been developing rapidly. The first is
cosmological simulations of the formation of the first stars and of 
the reionisation period. The second is the observations of extremely 
metal poor stars. 
In this paper, we present pre--supernova evolution models of massive 
rotating stars at very low $Z$ ($Z=10^{-8}$) and at $Z=0$. Rotation has a
strong impact on mass loss and nucleosynthesis. Models reaching break--up
velocities lose up to ten percents of their initial mass.
In very low $Z$ models, rotational and convective mixing enhances
significantly the surface content in carbon, nitrogen and oxygen (CNO)
when the star becomes a red supergiant. This induces a strong mass
loss for stars more massive than about 60 $M_\odot$. Our models predict type
Ib,c supernovae and gamma--ray bursts at very low $Z$. Rotational mixing
also induces a large production of CNO elements, in particular of
primary nitrogen. The stellar wind chemical composition is compatible with the
most metal--poor star know to date, HE 1327--2326, for CNO elements. Our
models reproduce the early evolution of nitrogen in the Milky Way.
\end{abstract}
\maketitle

\section{Physics and models at very low metallicities}
What is the difference between very low $Z$ stars compared to 
solar metallicity stars? Lower $Z$ stars are more compact due to lower opacities. 
For example, the radius of a 20 $M_\odot$ star at $Z=10^{-8}$ is about 
1/4 of a 20 $M_\odot$ star at solar metallicity. It is also more difficult for
very low $Z$ stars to reach the red supergiant (RSG) stage.

For metal--free stars ($Z \lesssim 10^{-10}$), the CNO cycle cannot
operate at the start of H--burning.  At the end of its formation, the 
star contracts until it starts He--burning because the
pp--chains cannot balance the effect of the gravitational force. Once
enough carbon and oxygen are produced, the CNO cycle can operate and the
star behaves like stars with $Z>10^{-10}$ for the rest of the MS. 
Shell H--burning still differs between $Z>10^{-10}$ and metal--free 
stars. 

Due to fragmentation and cooling properties of metal--free gas, the first
 stars formed are thought to have been more massive than solar metallicity 
stars but by how much and for how long is still debated 
\cite{BL04,SOIF06,SL06}. In particular, the nucleosynthetic products of 
pair--instability SNe are not seen at the surface of extremely metal 
poor stars observed in the halo of our galaxy \cite{CL04,HW02,UN05}. 

Since mass loss depends on metallicity
 one expects to have much weaker winds at
very low $Z$. However, the important point is to know which 
elements contribute the most to opacities and mass loss. Are these the 
iron group elements or light elements like CNO? In this work, it was 
assumed that all elements have the same importance. Theoretical studies 
have started answering this question \cite{VKL01,VdK05,VL05} and this 
topic will be studied further in a future work.

The stellar evolution code used to calculate the stellar models
 is described in detail in Hirschi \etal \ \cite*{psn04}.
Convective stability is determined by the 
Schwarzschild criterion. 
Convection is no longer treated as an instantaneous mixing but as a 
diffusive process from oxygen burning onwards.
The overshooting parameter is 0.1--0.2 H$_{\rm{P}}$ 
for H-- and He--burning cores 
and 0 otherwise.  
The reaction rates are taken from the
NACRE \cite{NACRE} compilation.
The mass loss is proportional to the square root of the metallicity. 
It is also dependent on the surface rotation velocity.
The centrifugal force is included in the structure equations.
The processes taken into account, which induce transport and mixing of 
angular momentum and matter, are meridional circulation and dynamical 
and secular shears.

Stellar evolution models studying the metallicity dependence were 
calculated \cite{H06,MEM06}. In addition, two grids of models were calculated, 
one for metal--free stars \cite{EMM06} and one for very low
$Z$ stars at $Z=10^{-8}$, which corresponds to the second stellar 
generation \cite{H06}. The initial velocity of the models range between 
500 and 800 $\rm km\,s^{-1}$ and were chosen such that the
angular momentum (and ratio of the initial velocity to the critical 
velocity) of the models are similar to solar metallicity models with 
an initial velocity of 300 $\rm km\,s^{-1}$. The fact that stars are much 
more compact at very low $Z$ explains why initial surface 
velocities are higher at very low $Z$. The choice of velocity is supported by the
fact that our models can reproduce the evolution of the N/O ratio (See Sect.
3.2).

\section{Evolution at very low metallicities}
\begin{figure}[!tbp]
\centering
\includegraphics[width=6cm]{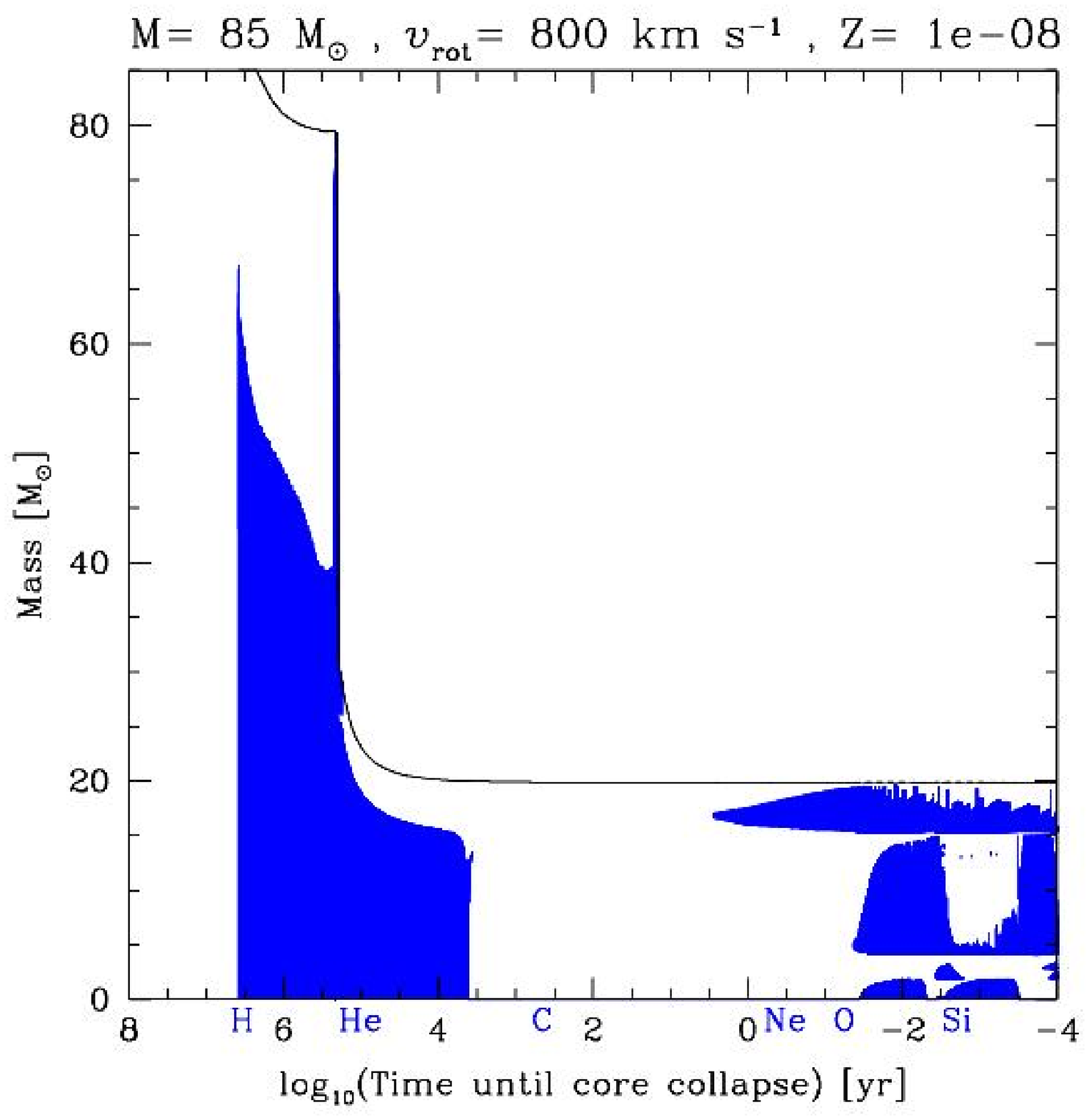}\includegraphics[width=6cm]{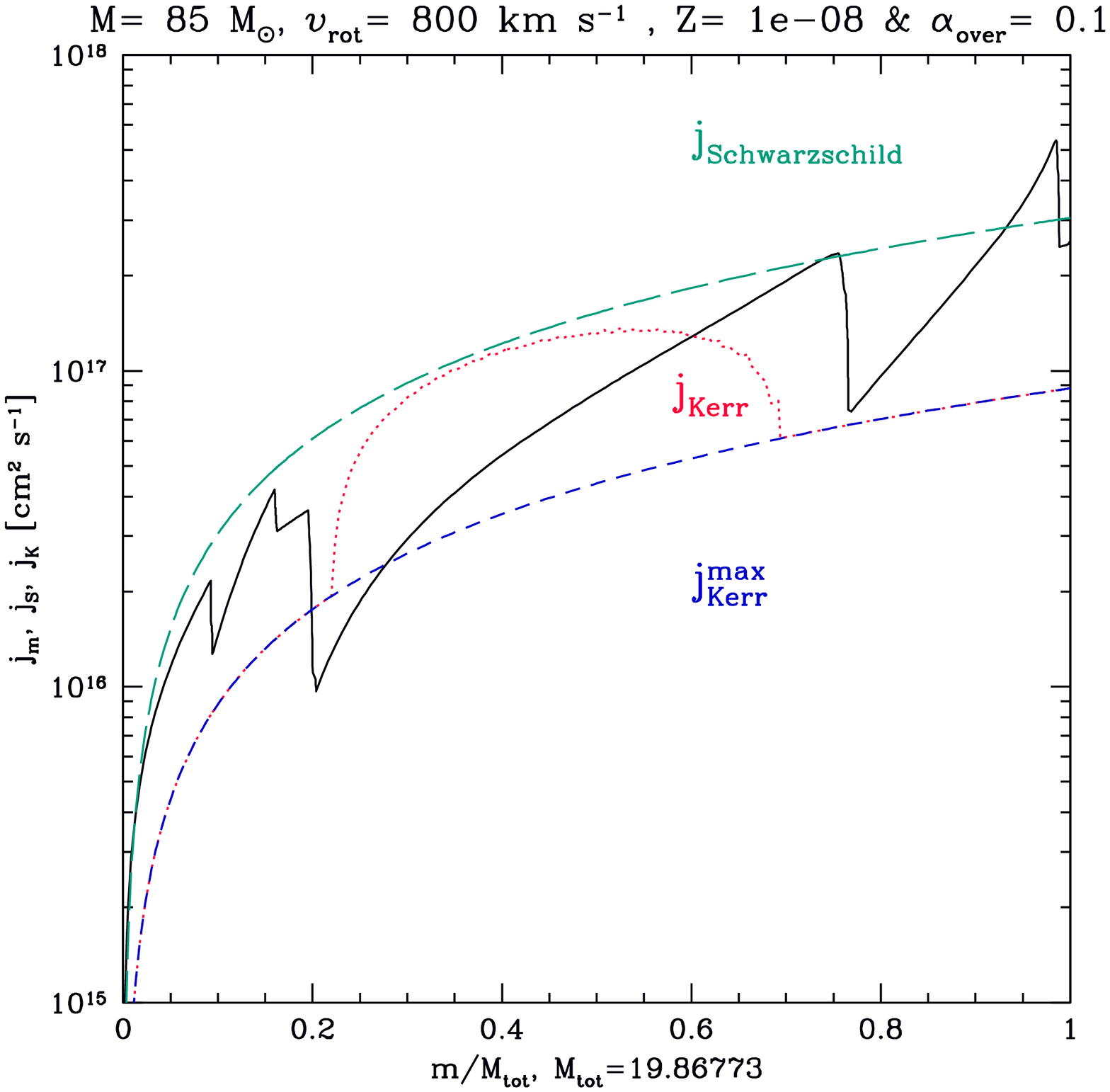}
\caption{{\it Left}: Stellar structure (Kippenhahn) diagram of the 85 $M_\odot$ 
model at $Z=10^{-8}$. The top solid lines represent the total mass, which decreases strongly when the star
becomes a red supergiant. The colored areas represent convective cores and
shells.
{\it Right}: Profile of the specific angular momentum at the pre--SN stage (solid line). 
The red dotted line shows the minimum angular momentum necessary in
order to form an accretion disk around a rotating black hole.
The blue short dashed and green long dashed lines show the minimum angular momentum
necessary for a maximally rotating and a non--rotating black hole,
respectively. This model has enough angular momentum in its core to form an accretion disk and therefore
may produce a GRB.
}
\label{m85}
\end{figure}

The standard picture is that very low $Z$ and metal--free stars do not 
lose any mass before they collapse. Rotation however changes this 
picture. First, very low $Z$ stars reach break--up velocities 
(centrifugal force added to the radiative force balancing the 
gravitational force) during the main sequence. Second,
rotational and convective mixing enriches the surface in primary CNO 
elements and therefore enhances mass loss.
Metal--free models reach break--up velocities and lose between 1 
(25 $M_\odot$) and 10 percents (200 $M_\odot$) of their initial mass. 
Note that this fraction may be higher in more massive stars.
In metal--free models, the surface does not become enriched 
significantly in CNO elements before the end of He--burning and the 
second effect cited above is not important. However, models at 
$Z=10^{-8}$ experience a strong enrichment in primary CNO and lose a 
large fraction of their initial mass for masses larger than about 
60 $M_\odot$. The surface enrichment starts when the model enters the RSG
stage and the surface metallicity becomes almost solar for CNO elements. 
Since we considered that all elements are important for mass loss, 
the mass loss is very high during the rest of the RSG stage.
Figure \ref{m85} ({\it left}) shows the structure evolution of the 
85 $M_\odot$ model. The top solid line represents the total mass of the model. 
The model loses about 10 percents of its initial mass due to the
break-up phenomenon during H--burning and then more than half of its 
initial mass during 
the RSG stage. This model produces a WO--type WR star and 
retains enough angular momentum in its core (Fig. \ref{m85} {\it right}) 
to produce a gamma--ray burst (GRB) via the collapsar model. Therefore, 
from the second stellar generation, our models predict important mass 
loss 
and SNIc--GRB events. This is in stark contrast with the standard picture where 
this model would not eject any of its matter during its lifetime or 
when it collapses to a black hole.

\section{Nucleosynthesis}
Rotationally induced mixing has an important impact on
nucleosynthesis. Carbon and oxygen produced into the He--burning core are
mixed in the H--burning shell and large amounts of primary nitrogen are
produced. At very low $Z$, rotationally induced mixing may increase the
nitrogen production by a factor 1000 or more. Part of the nitrogen is
transformed into $^{22}$Ne and other intermediate mass elements via 
double alpha capture during shell He--burning. $^{22}$Ne is a neutron
source for s--process in massive stars. In metal--free stars, s-process
is inefficient because it has to start from carbon. However, in very
low $Z$ stars, small amounts of iron group elements are present and
s--process may produce interesting results since there is a high neutron to
seed ratio. We will calculate s--process
nucleosynthesis in future studies. In this section, we compare our models to
observations of extremely metal poor (EMP) stars.

\subsection{Carbon--rich EMP stars}
\begin{figure}[!tbp]
\centering
\includegraphics[width=6cm]{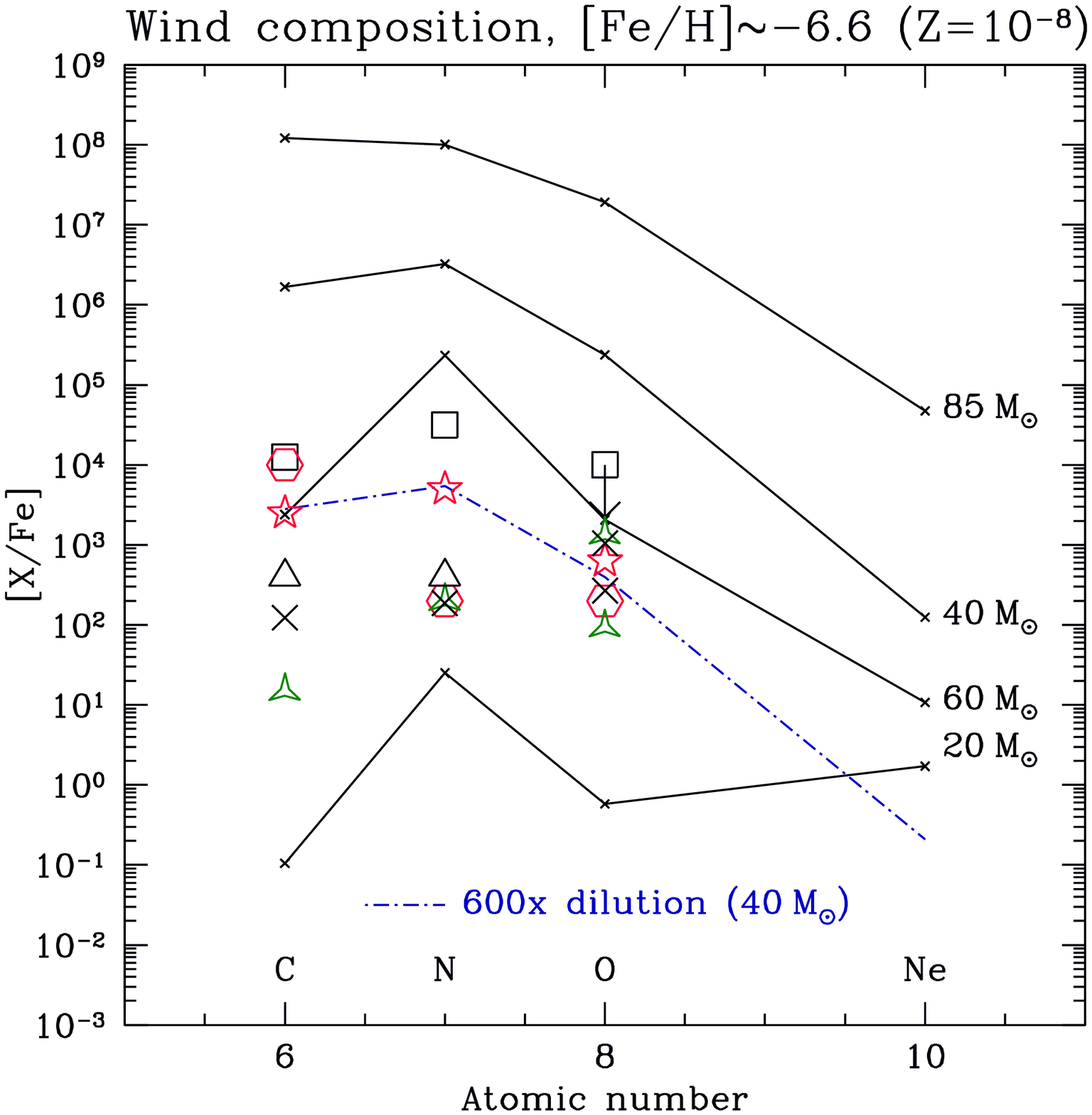}\includegraphics[width=6cm]{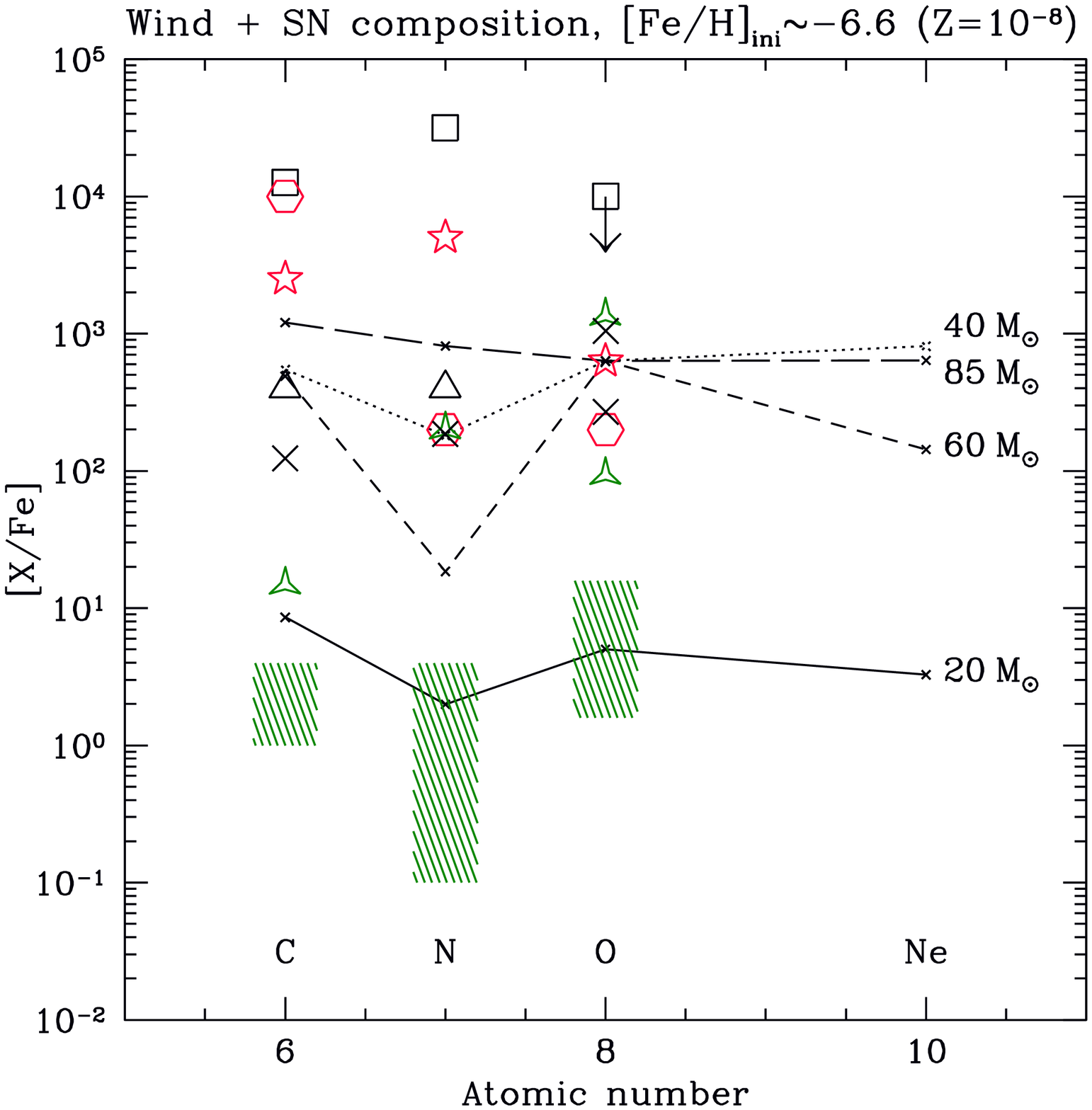}
\caption{Composition in [X/Fe] of the stellar wind ({\it left}) and the
sum of the wind and SN ejecta ({\it right}) for the $Z=10^{-8}$ models.
The lines represent predictions from the models.
The {\it empty triangles} and the {\it red stars} correspond to the
observed surface abundances of G77-61 
(Plez and Cohen, 2005) 
and to
the new (3D/NLTE corrected) estimates for HE1327-2326 
(Frebel et al., 2006), 
respectively. These are both unevolved stars. The abundances
of HE1327-2326 are best reproduced by the wind composition of the 40
$M_\odot$, diluted by a factor 600 with pristine gas. 
This figures are taken from Hirschi 
(2006).
}
\label{cemp}
\end{figure}
About one quarter of EMP stars are carbon rich. One quarter of the
carbon--rich EMP stars are thought to have been enriched by a weak s--process
probably taking place in massive stars
\cite{RANB05}. The most metal poor star known to date ([Fe/H]$\sim$-5.5), 
HE1327-2326, belongs to this last group. It is thought that its material 
has been enriched by only one or a few massive stars. It is therefore
possible to compare directly the chemical output of our models to such a
star. Figure \ref{cemp} shows that the best fit for this star is obtained
by diluting the composition of the 40 $M_\odot$ models at $Z=10^{-8}$ 
by a factor 600 within interstellar medium (ISM) gas.
In this scenario, HE1327-2326 is a third generation star. First, a
metal--free star polluted the ISM to very low $Z$. Then a second
generation star, like the 40 $M_\odot$ model at $Z=10^{-8}$, polluted 
(only through its stellar wind) the ISM, out of which HE1327-2326 formed.

\subsection{Primary nitrogen}
\begin{figure}
\centering
\includegraphics[width=8cm,angle=0]{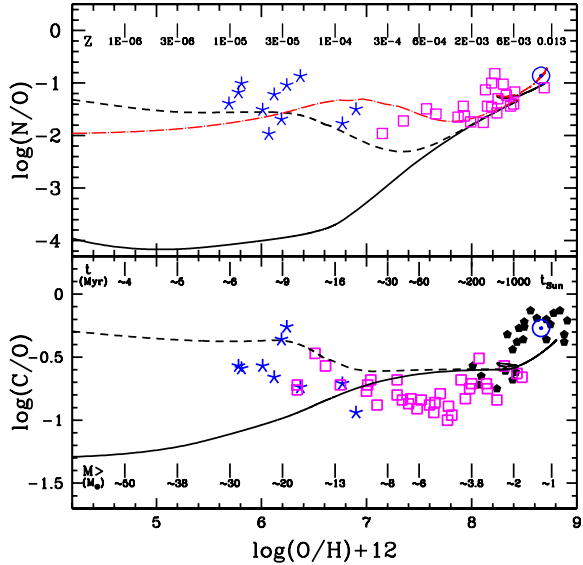}
\caption{{\it Upper panel}: Solar vicinity diagram log(N/O)
  vs. log(O/H)+12. 
{\it Lower panel}: Solar vicinity diagram log(C/O)
vs. log(O/H)+12. The symbols are observations 
(Spite et al., 2005; Israelian et al., 2004; Akerman et al., 2004; Nissen, 2003). 
The
black dashed line represents the chemical evolution model including the 
very low $Z$ stellar yields. It fits both the evolution of N/O and C/O.
This figure is taken from 
Chiappini \etal \ (2006).
}
\label{cno}
\end{figure}
The stellar yields of the very low $Z$ stars were used in a galactic
chemical evolution model \cite{CH06}. The evolution of CNO elements is
shown in Fig. \ref{cno}. The previous models (solid black line) clearly
underproduced nitrogen at very low metallicities. Using our models at
$Z=10^{-8}$, the evolution of nitrogen for log(O/H)+12 $<$ 7 (or [Fe/H]$<$ $-$3) 
is much better reproduced (black
dashed line). Our models also predicted an increase of the C/O ratio
towards very low metallicities (Fig. \ref{cno} {\it lower panel}), which also
fits the observations. 
These results give further support to the idea that stars at very low
 metallicities had high surface rotational velocities, of the order of 
500-800\,km\,s$^{-1}$.


\end{document}